\documentclass[aps,pre,twocolumn]{revtex4}
\usepackage{amsmath}
\usepackage{amssymb}
\usepackage{color}
\usepackage{graphicx}
\usepackage{indentfirst}

\newcommand{\sFrac}[2]{{\textstyle\frac{#1}{#2}}}
\newcommand{\pair}{{(\hskip-0.03cm\alpha\hskip-0.03cm)}}
\newcommand{\generalPair}{{ij}}

\newcommand{\be}{\begin{equation}}
\newcommand{\ee}{\end{equation}}
\newcommand{\bea}{\begin{eqnarray}}
\newcommand{\eea}{\end{eqnarray}}
\newcommand{\ba}{\begin{eqnarray}}
\newcommand{\ea}{\end{eqnarray}}

\date{\today}
\begin{document}
\title{Marginal Stability in Structural, Spin and Electron Glasses}
\author{Markus M\"uller and Matthieu Wyart}
\affiliation{${\ }^1$The Abdus Salam International Center for Theoretical Physics, Strada Costiera 11, 34151 Trieste, Italy. Email: markusm@ictp.it\\
${\ }^2$Center for Soft Matter Research, Department of Physics, New York University, New York, NY 10003. Email: mw135@nyu.edu
}

\date{\today}
\begin{abstract}
We revisit the concept of marginal stability in glasses, and determine its range of applicability in the context of avalanche-type response to slow external driving. We argue that there is an intimate connection between a pseudo-gap in the distribution of local fields and crackling in systems with long-range interactions. We show how the principle of marginal stability offers a unifying perspective on the phenomenology of systems as diverse as spin and electron glasses, hard spheres, pinned elastic interfaces and the plasticity of soft amorphous solids.  
\end{abstract}
\maketitle
\section{Introduction}
In glassy materials the dynamics is so slow that thermal equilibrium cannot be reached. In these systems, states of equal energy are  not equiprobable. In order to describe their physical properties, one thus faces the challenge of understanding the ensemble of configurations visited by the dynamics, a problem central to many complex systems. This ensemble can in principle be computed from an accurate description of the dynamics, but in general this task is very difficult. 

Here we  review the principle of marginal stability, which provides guidance about the ensemble of configuration explored in a variety of glasses, and controls some of their key physical properties. This concept has been introduced in disordered insulators,  the so-called electron glasses ~\cite{Pollak, efros,LeeDavies, PollakBook,monroe,eglass, MasseyLee, goethe,DobroPankov,markus, MarkusPankov, PalassiniGoethe}. It has also been applied to long-range spin glasses \cite{thouless,Pazmandi,moore,horner,MarkusEPL,MarkusSKavalanchesPRB, XYglass}, and, more recently, to the simplest structural glasses, namely packings of hard particles \cite{Wyart12,Lerner13a,DeGiuli14,Kallus13,Kallus14}. It thus offers a starting point to compare these glasses, and to describe them within a common framework. 

\begin{figure}[ht]
\includegraphics[width=0.4 \textwidth] 
{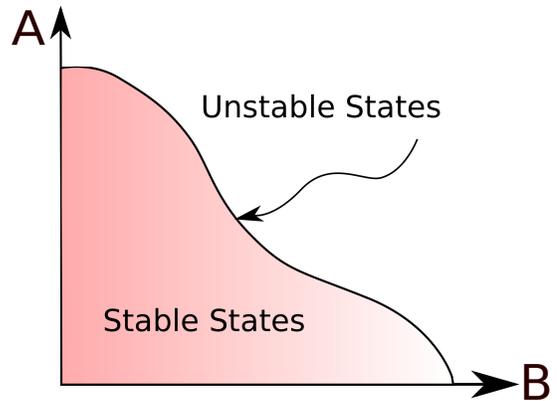}
\caption{Schematic  stability diagram in configuration space. $A$ and $B$ are observables characterizing the configurations visited.  The blue line corresponds to marginal stability: it separates regions where  excitations are stable and unstable, respectively. The red arrow illustrates a dynamical trajectory of a system which is cooling from an initial high temperature phase. The system is initially unstable, until it reaches the marginality line. At this point the excitations become stable. If these excitations are the main drive of the dynamics, the latter will slow down very rapidly as soon as the system enters into the stable region, and the system will freeze very close to the marginal stability line. For an infinitely rapid quench, in the absence of inertia, the system could lie on the marginality line.  On the stability line soft excitations are abundant, and rich dynamics, such as crackling noise, can occur.
}
\label{fig:marginal}
\end{figure}

The concept of marginal stability is illustrated in  Fig.~\ref{fig:marginal}. It is based on the identification of 
elementary excitations and their stability. The diagram shows three regions of configuration space: a region where excitations are absolutely stable, an unstable region, and  a marginal manifold  separating them. 
In certain mean-field models of glasses, marginality can be shown to be present at equilibrium in the low-temperature phase ~\cite{markus,parisi,Charbonneau14}.  However, in finite dimensions, marginal stability comes about more naturally in {\em dynamics} and in out-of-equilibrium situations.   Indeed, assume that the considered elementary excitations  are unstable at high temperature. Instead they must be stable at zero temperature, if equilibrium were achievable. Consider the dynamics  following a very rapid quench from high temperature. As it progresses, the system eventually reaches the marginal manifold, where excitations stabilize. If the relaxation of these excitations are a key drive of the dynamics, one expects that the system dramatically slows down at that point, falls out-of-equilibrium and freezes near the marginal manifold. While the simplicity of this argument is appealing, its domain of applicability has remained  unclear. In this review we aim at a classification of random systems to which it does apply.

When it applies, marginal stability has important consequences, both because it reduces the configuration space to be considered, and because it implies an abundance of soft excitations that can strongly affect the response of the system. 
Sometimes marginal excitations are continuous, and are associated with the lowest eigenvalues  of the Hessian matrix of an appropriate energy functional. This situation occurs for the vibrational modes  of some amorphous solids, such as compressed soft particles~\cite{Wyart05b,DeGiuli14}, colloidal glasses \cite{Brito06,Brito09,DeGiuli14b} or Lennard-Jones systems~\cite{Marruzzo13a}, in soft p-spin systems~\cite{CuKu, Kurchan}, or vector spin glasses~\cite{XYglass}.
In this review we shall instead focus on discrete excitations, including spin flips in Ising spin glasses, the motion of electrons between localized states in Coulomb glasses, and the opening and closing of contacts in packings of hard spheres. As we shall recall, stability in these systems implies that the density of excitations must vanish at zero energy, i.e. the presence of a pseudo-gap. At low enough temperature, when these  systems are driven (by increasing the magnetic field in a spin glass, the gate voltage in the Coulomb glass, or the anisotropy of stress in packings), numerical simulations often report that the dynamics is bursty, and occurs via rearrangements in which many excitations are involved. The size of those "avalanches"  is usually power-law distributed. They produce what is often referred to as crackling noise~\cite{Sethna01}. The relationship between crackling noise and the presence of a pseudo-gap has remained unclear. In this work we shall argue that there is a one-to-one correspondence between these two phenomena under certain conditions of driving. 

Crackling noise also occurs in a distinct class of systems, that  display a so-called self-organized criticality~\cite{Bak87}. A typical example is the depinning transition of an elastic line in a disordered environment: at some critical force $F_c$ the line motion is jerky and consists of broadly-distributed avalanches. Because the velocity of the line under some linear force $F$ follows a critical power law $v=v_{\rm typ} (F/F_c-1)^\beta$ where $v_{\rm typ}$ is a typical velocity, the line spontaneously sits near the critical point if it is driven at a slow pace $v\ll v_{\rm typ}$ \cite{Fisher98}. This situation is  different from the marginal stability that follows a rapid quench reviewed here, which, as we shall see, results in a critical {\em phase} that exists for a range of fields, instead of at a single value. In our classification below, the depinning transition of a manifold with short-range elasticity is not considered marginal, as crackling does not occur if the external force changes, as long as $F<F_c$.

Our manuscript is organized as follows. In Secs.~\ref{spin}, \ref{electron} and \ref{hs}, we introduce spin, electron and hard sphere glasses, respectively. In each case, we present  stability arguments leading to an upper bound on the density of excitations at low-energy. For spin glasses, we propose a novel  argument which corrects previous, incorrect considerations. In these sections we also review the evidence for the saturation of the stability bounds,  the effects of the associated pseudo-gaps on physical properties, and the presence of crackling noise under external driving. In Sec.~\ref{framework} we  propose a classification of marginally stable systems. We show that under broad conditions, if the system is driven in the absence of thermal noise by changing the field on some finite interval,  any dynamics that relax excitations one by one must display crackling if a pseudo-gap is present. We then argue  that for interactions that do not decay with distance (as occurs effectively in packing of hard spheres) such crackling is possible only if the pseudo-gap saturates the stability bound. From this we conclude that stability is generically marginal under these conditions. Our argument also implies that crackling must occur if the interactions are frustrated and sufficiently long range,
and that it does not disappear if local interactions are added, even if the latter are much stronger than the long-range interactions. On the other hand, if interactions are purely short-range, our argument suggests that the system cannot crackle in an interval of fields. Thus long-range interactions are relevant, as far as the dynamics under driving conditions is concerned. Finally, we  discuss how other situations fit into our classification. In particular, we discuss the depinning of elastic manifolds in random environment, and the plasticity that occurs in soft amorphous solids under shear. For the latter, we predict that crackling must occur even below the yield stress where continuous flow is maintained, in agreement with recent numerical studies.


\section{Spin glasses}\label{spin}
Spin glasses are magnets with random exchange Hamiltonians   
\bea
H = -\frac{1}{2}\sum_{i\neq j} s_i J_{ij} s_j,
\label{HamSG}
\eea      
where $J_{ij}$ are frustrated exchange interactions between the local spins $s_i$. Due to the frustration no obvious ordering pattern exists at low temperature. Below we will mostly discuss Ising systems with $s_i=\pm 1$. 
Upon a quench into the glass phase, such frustrated magnets tend to become trapped into metastable states with apparently random magnetization patterns. They exhibit very slow relaxation dynamics, which does not reach equilibrium on experimental timescales.     

Spin glasses come in a large variety of microscopic realizations of Hamiltonians of the form of Eq.~(\ref{HamSG}). The principle of marginal stability is particularly fruitful in  glasses in which the exchange interactions are long range in the sense that they decay in space as a power law $J_{ij}\sim 1/|\vec{r}_i-\vec{r}_j|^\gamma$ with $\gamma\leq d$, apart from their random sign. This is the case, e.g., for metallic spin glasses where the spins are coupled via RKKY interactions $|J_{ij}|\sim 1/r_{ij}^d$.
An extreme limit is given by the Sherrington-Kirkpatrick (SK) model~\cite{SK}, where in Eq.~(\ref{HamSG}) every spin interacts with every other one via a random exchange $J_{ij}$, taken to be a Gaussian variable of zero mean and variance $1/N$, $N$ being the total number of spins. We will also analyze the case of short range spin glasses of connectivity $Z$ and nearest-neighbor couplings with zero mean and variance 
\bea
\label{couplingscaling}
\overline{J_{ij}^2}=1/Z. 
\eea
The latter guarantees a well-defined limit for $Z\to \infty$. Interestingly, while the thermodynamics is expected to approach the limit of the SK model with $N=Z\to \infty$, we will argue that the dynamics of the two models is rather different. Below we will discuss especially the Edwards-Anderson model, where the spins sit on a $d$-dimensional hypercubic lattice with $Z=2d$.  

A key observable in low $T$ configurations is the distribution of local fields, 
\bea
P(h)= N^{-1}\sum_i\delta(h-h_i),
\eea
where $h_i = \sum_{j\neq i} J_{ij} s_j$ is the local exchange field acting on spin $i$. 
Glasses with long-range interactions exhibit two connected  features, both related with marginal stability~\cite{dobro03,Pazmandi,moore, horner,MarkusEPL,MarkusSKavalanchesPRB,Katzgraber}: { (i)} a pseudogap in the field distribution $P(h)$; and { (ii)} 
a scale free distribution of magnetization bursts observed during slow magnetization processes, e.g., along a pseudo-adiabatic hysteresis loop. In random magnets, such bursts or avalanches are the analogue of Barkhausen noise, which is well-known in ferromagnets.~\cite{barkhausen,Sethna01}
A remarkable aspect of long-range interacting glasses is the fact that the distribution of avalanche sizes is scale free all along the hysteresis loop~\cite{Pazmandi}, in contrast to what one finds in many ferromagnetic model systems, such as the random field Ising model, where such criticality requires fine-tuning, both of the disorder strength and the externally applied field.~\cite{AvalanchesRFIM,vives}

\subsection{Distribution of local fields in mean-field spin glasses}
The shape of the distribution $P(h)$ underlies certain natural constraints. The scaling (\ref{couplingscaling}) ensures that typical local fields remain $O(1)$ in the limit of large $Z$. At high temperature, $P(h)$ is a featureless  Gaussian. However, upon a rapid quench  to $T=0$, unstable spins (with orientation opposite to their local field) are flipped, which on average  stabilizes the spins that they are coupled to. Thus, the local fields have tendency to become larger, and a pseudo-gap  starts to form at $h< 1$. In a locally stable state all spins are aligned with their local field.
Since there is no further scale in the problem, for large enough $Z$, one expects the distribution at low fields to be given by a simple power law 
\bea
\label{pseudogap}
P(J_{\rm typ} \lesssim h\lesssim 1)= C h^\theta,
\eea
which levels off only for fields below the typical coupling strength between neighbors, $h\lesssim J_{\rm typ}\equiv 1/\sqrt{Z}$.

Stability with respect to elementary spin flips imposes constraints on the pseudo-gap exponent $\theta$. Assuming that the local fields are uncorrelated, one finds that the smallest field (in absolute value) among $Z$ neighbors of a given site is of order $h_{\rm min}\sim (ZJ_{\rm typ})^{-1}\sim Z^{\theta/2-1}$ for $\theta\leq 1$.
Now consider sites $0$ where $h_0$ is itself of order $ h_{\rm min}$ or smaller, and its 
softest neighboring site, call it $s$. It is reasonable to assume that with finite probability, their interaction is unfrustrated. (In the SK model discussed below, one can argue rigorously that there must be spins among the $O(\sqrt{\log{N}})$ softest ones whose coupling is unfrustrated. However, we expect the fraction of unfrustrated pairs to be finite.) 
The simultaneous flip of two such spins $s_0, s_s$ costs an energy
\bea
\label{twospinflip}
\Delta E_{0s} = 2(|h_0|+|h_s|)- 2|J_{0s}|.
\eea
In order for the last term not to compensate the first  ones with very large probability, one must have $h_{\rm min}\gtrsim J_{\rm typ}$, and thus 
\bea
\label{spinglassbound}
\theta \geq 1. 
\eea
In numerics one observes the scaling $P(h=0) \sim 1/Z^{1/2}$ of the pseudo-gap, implying that the constraint is just marginally satisfied with $\theta=1$~\cite{localfields,Katzgraber}. In Sec.~\ref{framework} we will explain  this observations dynamically.

An interesting limiting case, which is amenable to rigorous thermodynamic calculations, is the fully connected SK spin glass, which corresponds to Eq.~(\ref{HamSG}) with Gaussian $J_{ij}$ of variance $1/N$ between all pairs of spins. In this case it is interesting to extend the stability argument to multiple spin excitations. 
Consider flipping a set  $\cal F$ of spins, forming a finite fraction of the $n\ll N$ spins in the smallest local fields. The associated energy change is
\bea
\label{SKcollective}
\Delta E= 2\sum_{i \in \cal F} |h_i|- 2\sum_{i, j\in \cal F}  J_{ij}s_is_j.
\eea
The first term is positive, and scales as $n(n/N)^{1/(1+\theta)}$. At first sight the second term might be expected to be negative. This was argued in Ref.~\cite{PalmerPond,AndersonLesHouches}, assuming that the spins in low fields tend to minimize their contribution to the full energy (\ref{HamSG}). However, taking a dynamical perspective it becomes clear that spins in low fields  are biased to be mutually frustrated: When a spin flips to relax its energy, it stabilizes those spins which it is not frustrated with. The spins whose stability is lowered are those that are frustrated with the majority of other soft spins that have relaxed previously. This observation is confirmed by numerical simulations~\cite{LeUs}, and implies that on average  the second term in Eq.~(\ref{SKcollective}) is positive. One can show that for $\theta=1$ it scales like $n^{3/2}/N^{1/2}$, as the first term. However, to assess stability one must consider {\em optimized}
sets ${\cal F}$ of $n$ spins (drawn from the softest $c\cdot n$ ones, with fixed $c=O(1)$), and thus study rare negative fluctuations of the second term in Eq.~(\ref{SKcollective}). Its standard deviation scales as $\sigma \sim  n/\sqrt{N}$. In the spirit of a random energy model, assume that the exchange sums for the $\exp[O(n)]$ different choices of ${\cal F}$ are independent and Gaussian distributed with variance $\sigma^2$. Typically the strongest negative fluctuations $f$ will be given by $\exp[O(n)] \exp[-f^2/2\sigma^2]\approx 1$, i.e., $f\sim -n^{3/2}/N^{1/2}$. For these fluctuations not to outweigh the first term and the average of the second one, it is necessary to impose $\theta\geq 1$. This consideration confirms that the pair constraint (\ref{spinglassbound}) is already sufficient to determine the marginal gap exponent. 

 Interestingly, for the SK model, saturation of the stability bound $\theta=1$ can be proven analytically  for the {\em ground state}~\cite{SommersDupont,Pankov}. The saturation is found to be closely connected to the thermodynamic criticality of Parisi's replica solution, which is at the verge of instability throughout the glass phase. We discuss this further in Sec.~\ref{framework}.

\subsection{Avalanches in spin glasses}
Avalanches in Ising spin glasses have been studied numerically along pseudo-adiabatic hysteresis loops where single spin flip updates in a $T=0$ dynamics was considered, both in the SK model~\cite{Pazmandi} and in short range  models of finite connectivity in finite dimensions and on random lattices~\cite{Katzgraber}. 
Spin glasses with continuous XY symmetry have been studied in the fully connected limit in Ref.~\cite{XYglass}.  

The analysis of Ising systems revealed an interesting phenomenology: Long range models exhibit a large number of mesoscopic avalanches, triggered by the slow increase of the external field. The distribution of their sizes $S$ (as measured by the number of flipped spins) was found to be power law distributed, with density 
\bea
\rho(S) \sim \frac{1}{S^\tau},\quad S< S^*,
\eea
where $\tau =1$ in the SK model ($\tau=3/2$ for electron glasses see below), and $S^*$ diverges with the system size. 
Instead, for short range interacting spin glasses, no such criticality was found~\cite{vives, Katzgraber}, despite the fact that the {\em static} equilibrium response in small fields is theoretically expected to exhibit power law response, both in spin glasses~\cite{MarkusEPL, MarkusSKavalanchesPRB} and random ferromagnets~\cite{MonthusGarel}. 

This dichotomy of the driven response shows that the range of the interactions has a crucial effect on the configurations visited in the dynamics: Long range interactions  tend to keep the systems in a critical, marginally stable state, while short range interactions allow the glass to settle with high probability into genuinely stable configurations. When rendered unstable the triggered rearrangements remain local there, while long-range interacting systems  generate scale-free avalanches. A theoretical explanation of these observations will be given in Sec.~\ref{framework} below.

Not only the states on the hysteresis curve, but also the equilibrium states undergo jerky jumps as the external field is increased. In the case of mean-field spin glasses, the statistics of those equilibrium avalanches can be computed analytically using replica methods.~\cite{MarkusEPL, MarkusSKavalanchesPRB}. 
Interestingly, one finds power law equilibrium avalanches with essentially the same characteristics as seen in the avalanches on the hysteresis loop: An avalanche exponent $\tau=1$ and a cut-off $S^*\sim N$ diverging with system size.

\section{Electron glasses}
\label{electron}

Electron or Coulomb glasses~\cite{Pollak,efros,LeeDavies, PollakBook} are long-range interacting frustrated systems which arise in strongly disordered systems such as semiconductors or granular metals. In these systems, electrons are localized and interact with each other via unscreened, long-range Coulomb interactions~\cite{eglass}. The competition between the disorder potential, which favors a random electron distribution, and the Coulomb repulsion, which favors uniform charge density, induces frustration. 
As a  representative model consider doped semiconductors described by 
\bea
\label{ESmodel}
H=\sum_i\epsilon_i n_i +\frac{1}{2}\sum_{i\neq j} (n_i-\nu)\frac{e^2}{r_{ij}}(n_j-\nu),
\eea
where (spinless) electrons occupy a fraction $\nu$ of the localization sites $i$, subject to a disorder potential, and mutual Coulomb repulsion. Each site can only host one electron, $n_i\in \{0,1\}$. One defines effective on-site energies as $E_i=\epsilon_i+\sum_{j\neq i}  e^2 (n_j-\nu)/r_{ij}$, and the associated single-particle density of states, $\rho(E)=N^{-1}\sum_i\delta(E-E_i)$. In a metastable state, all sites with $E_i<\mu$ are occupied and all other sites empty, where $\mu$ is the chemical potential.

\subsection{Coulomb gap}
Similarly as in long-range spin glasses, stability with respect to single particle moves enforces a Coulomb pseudo-gap, $\rho(E)\sim E^\alpha$.
Efros and Shklovskii~\cite{efros} derived the constraint $\alpha\geq d-1$ considering the stability with respect to an electron hop from an occupied site $i$ to an empty one $f$ - the analogue of the pair of spin flips considered above. Moving the electron from $i$ to $f$ costs the energy
\bea
\label{EScondition}
\Delta E_{i\to f} = E_f-E_i -\frac{e^2}{r_{ij}} , 
\eea
which must be positive in any locally stable configuration. The term $E_f-E_i$ is always positive. In contrast to spin glasses, the last term is strictly negative and subtracts the particle-hole attraction between $i$ and $f$, which otherwise would be doubly counted in $E_i$ and $E_f$. To guarantee stability, low energy sites must have a suppressed density, otherwise there would be abundant pairs of sites which violate the constraint (\ref{EScondition}). Neglecting strong correlations between sites and assuming $\rho(E)\sim |E|^\alpha$, the smallest $E_i$ within a region of diameter $R$ scale as $E_{\rm min}(R) \sim R^{-d/(\alpha-1)}$. Taking the softest occupied and empty sites $i$ and $f$ in such a region, they can only offset the negative term $-e^2/r_{if}\sim 1/R$ at large distances, if 
\bea
\label{ESbound}
\alpha\geq d-1.
\eea
The marginally stable density of states, 
 \bea
 \rho(E) = A |E|^{d-1}
 \eea
 first proposed by Efros and Shklovskii~\cite{efros}, is indeed observed in typical metastable states which satisfy {\em only} the minimal two-particle stability criterium discussed above.~\cite{Palassiniunpublished}. 

However, when multiparticle stability constraints are enforced, the pseudo-gap might be suppressed beyond the bound $\alpha=d-1$, in particular in $d\geq 3$ dimensions~\cite{efros,Skinner}, where stability with respect to soft compact dipolar excitations has to be considered, on top of single charge excitations. Such a tendency is observed in numerical studies of the distribution of single site excitations~\cite{DrittlerMoebius, goethe}. However, it had long been argued~\cite{efros}, that the above bound should  truly be saturated only for {\em dressed} excitations (expected to be relevant for transport). Those are local charge insertions or removals with concomitant dipolar charge relaxations in their close neighborhood. This conjecture was based on the idea that stability constraints involving more than two sites are likely to concern sites within the polarization clouds of two main charge excitations, and are thus already contained in the stability constraint for one or two dressed excitations. However, this reasoning can hardly cover stability constraints between several sites that are more distant from each other than the typical diameter of the putative polarization clouds.   

The prediction of a saturated Efros-Shklovskii bound is also reached by a replica mean-field theory of the Coulomb glass~\cite{markus, DobroPankov,MarkusPankov}, in the framework of which one finds a thermodynamically marginal glass phase, and a density of states of dressed excitations, which saturates the Efros-Shklovskii bound on $\alpha$. Within mean-field theory the latter is closely connected to the thermodynamic criticality of the glass phase, similarly as in the case of the SK spin glass.~\cite{MarkusPankov}  A numerical verification of the saturation of the bound for dressed excitations has not been achieved, however.     

We caution, however, that even if for the relevant excitations the exponent $\alpha=d-1$ is universal, the close analogy with the SK model suggests that the prefactor $A$ is probably non-universal, but depends on the level of relaxation that has occurred to reach the considered metastable state. 

\begin{figure}[h]
\includegraphics[width=0.5\textwidth]{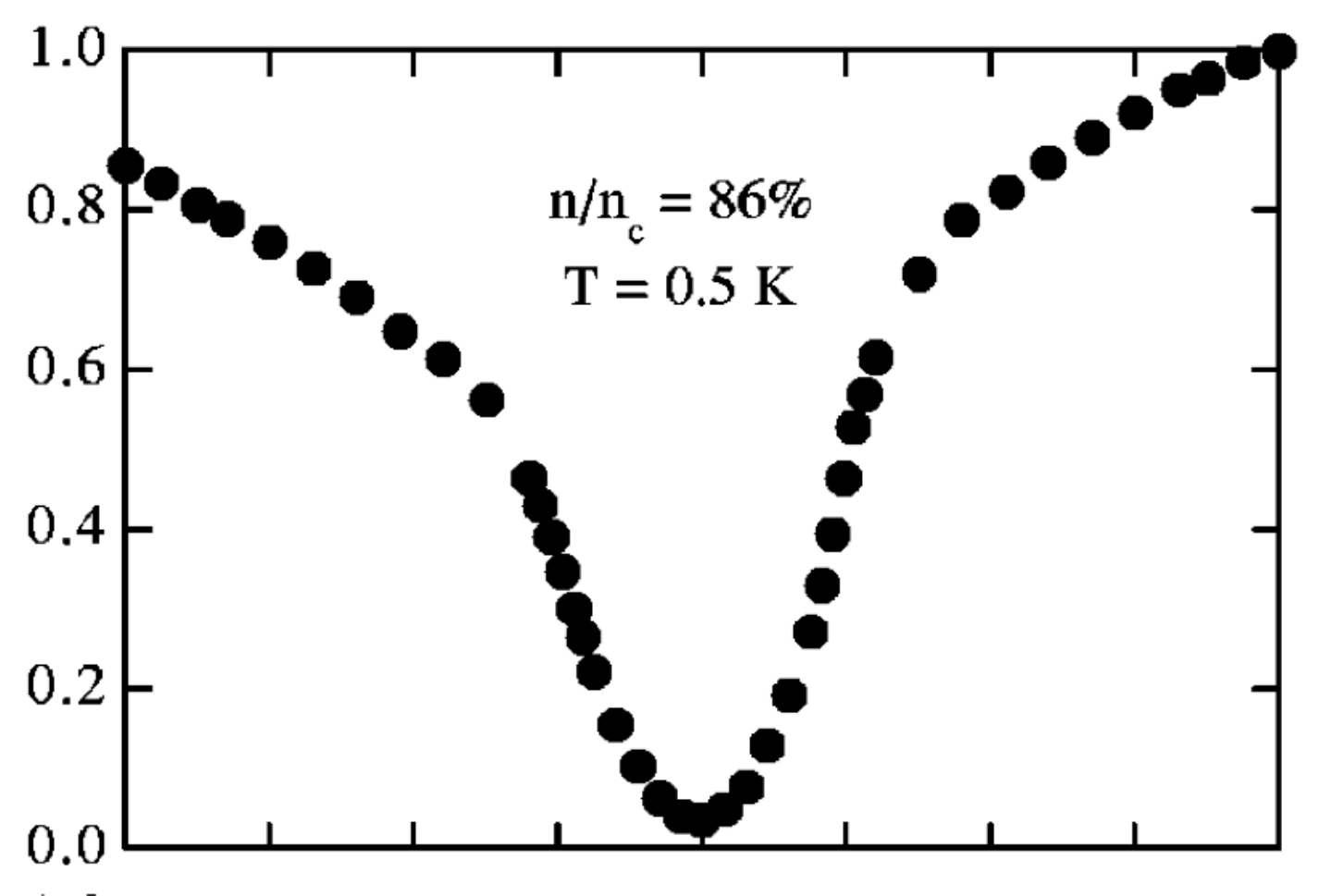}  
\caption{Stability in electron glasses with unscreened Coulomb interactions requires a pseudo-gap in the single particle density of states. The Efros-Shklovskii bound $\rho(E)\leq CE^{d-1}$ appears to be nearly saturated in tunneling experiments, as well as in numerical simulations. (From \cite{MasseyLee}).}
\label{fig:Cbgap}
\end{figure}

The Coulomb gap is easily seen in numerical simulations of the model (\ref{ESmodel}). In experiments, 
it can be detected in tunneling through a broad junction, cf.~Fig.~\ref{fig:Cbgap}, or indirectly via  measurements of d.c. resistance, which is sensitive to the density of single particle states at low energies. A pseudo-gap with exponent $\alpha$ leads to an enhanced variable range hopping resistance of the 
form $\rho(T)\sim \exp[(T_0/T)^{\frac{\alpha+1}{d+\alpha+1}}]$. 

The electronic equivalent of a slow magnetization process is the slow charging of the electron glass. A difference consists, however, in the fact that the extra charge has to enter the system from its contacts to a charge reservoir, whereas  spins can simply flip in response to an external field. Charging experiments are very challenging~\cite{monroe}, and so far there is no direct experimental evidence of avalanches in such systems. However, numerical simulations and the theoretical considerations presented in Sec.~\ref{framework} below suggest that avalanches of charge redistributions should occur at very low temperatures. However, their size and character depends on the allowed dynamical moves.~\cite{PalassiniGoethe,Andresen2} 

Arguments as for electron glasses apply to any frustrated system with discrete degrees of freedom and long-range interactions which decay as $1/r^{\gamma}$, where 
one obtains  the bound $\alpha\geq d/\gamma -1$. Examples are logarithmically interacting objects - such as vortices in 2d superconductors~\cite{Taeuber} and electrons in thin, highly polarizable films~\cite{LarkinKhmelnitskii} -  or dipolar systems with interactions $1/r^3$.~\cite{dipolar}

\subsection{Avalanches}
In electron glasses avalanches have been studied in response to local excitations, such as a charge or a local particle-hole excitation~\cite{PalassiniGoethe}. The latter were  found to induce scale free avalanches, distributed according to $P(S)\sim S^{-\tau}\exp[-S/S^*]$ with $\tau =3/2$ and $S^*\propto L$. However, this scale-free behavior was  observed only if electron hops in the avalanche were allowed to be arbitrarily long. In contrast, $S^*$ remained finite in the thermodynamic limit, if the distance of electron hops is restricted. The latter is in agreement with the findings of Ref.~\cite{Andresen2}, which studied the response via restricted hops, as triggered by an electric field.
These results were interpreted in terms of simple branching processes~\cite{PalassiniGoethe}, similar to the ones we will discuss in Sec.~\ref{SpinglassApp} for short range spin glasses. In that section we will rationalize the dependence on the admitted hopping distance.


\section{Sphere Packings} 
\label{hs}

The most common glasses are amorphous solids, the simplest example being frictionless hard spheres. 
The control parameter is the packing fraction  $\phi$. In three dimensions, if crystallization is avoided, spheres undergo 
a glass transition around $\phi_g\approx 0.58$. For $0.58<\phi<\phi_c\approx 0.64$, despite the particles' ability to rattle on small distances, an applied stress does not 
relax to zero on experimental time scales, and the material thus acts as a solid \cite{Pusey87,Mason95b,Parisi10}.  $\phi$ can be increased up to $\phi_c$ (depending on the protocol)  where particles make permanent contacts, and no rattling is possible. These states realize the random-close packing, used to model granular materials and emulsions. Here we do not discuss the case $\phi>\phi_c$ that can be achieved using soft particles  \cite{Liu10, Wyart05b, Hecke10, Alexander98,OHern03a}, where elastic properties and transport display scaling as $\phi_c$ is approached, see \cite{DeGiuli14,Lerner14} for recent discussions. 

\subsection{Structure at random close packing}

\begin{figure}[ht]
\includegraphics[scale = 0.43]{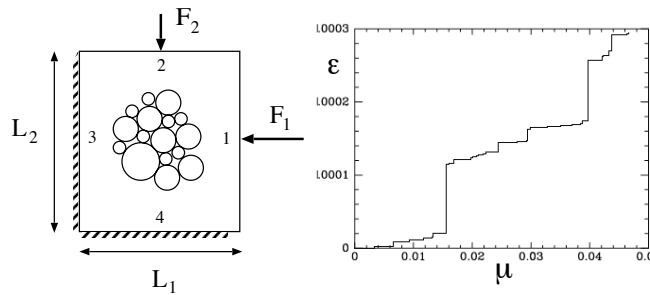}
\caption{Left: illustration of the box of initial length $L$ at some pressure $p$, corresponding to  forces of identical amplitude on each side $F_1=F_2=F$. Then $F_2$ is increased adiabatically leading to a growing stress anisotropy $\mu=F_2/F_1-1$. Particle rearrangements lead to a change of box shape, and to a strain $\epsilon=(L-L_2)/L$. Right: evolution of $\epsilon$ with $\mu$ showing crackling noise,  from \cite{Combe00}. }
\label{fig3}
\end{figure}

Consider a packing  of $N$ spheres of diameter $\sigma_0$ contained in a in $d$-dimensional hyper-cubic box of  volume $\Omega$ with rigid walls, cf. Fig.~\ref{fig3}. The walls apply a pressure $p$ that fixes the scale $\langle f \rangle \sim \sigma_0^2 p$ of contact forces.  ${\vec F}_i$ denotes the external force exerted by the wall on all the particles $i$ in contact with it,  
$\vec{R}_i$ is the position of particle $i$, $\vec{R}_\generalPair\equiv\vec{R}_j - \vec{R}_i$ and 
$r_\generalPair \equiv \sqrt{\vec{R}_\generalPair\cdot\vec{R}_\generalPair}$. 

{\it Coordination $z$:} 
Mechanical stability  requires that there exist no collective continuous motions of the particles during which particles will not overlap. If such a floppy mode existed, the system could flow along it. This  implies that the number of degrees of freedom $Nd$ is smaller than the  number of contacts $N_c$. Thus,  the average number of contacts per particles satisfies $z\equiv 2N_c/N\geq 2d\equiv z_c$, as derived by Maxwell \cite{Maxwell64}.
Rapidly compressed, or poly-disperse packings of hard frictionless spherical particles are in fact {\it isostatic} \cite{Alexander98}:  $z$ is \emph{just sufficient} to guarantee mechanical stability and $z=z_c$. Indeed if the number of constraints were larger, particles would generically overlap \cite{Alexander98,Tkachenko99,Moukarzel98}.


{\it Pair distribution function:} At $\phi_c$,  surprisingly, many particles are almost touching, but not quite. The distribution function $g(h)$ is the probability that two neighboring particles are separated by an interstice or gap $h\equiv r_{ij}-\sigma_0$. $g(h)$   has a singularity near contact: 
\be
\label{gap}
g(h)\sim \frac{\sigma_0^{\gamma-1}}{h^{\gamma}},
\ee
empirically $\gamma\approx 0.4$ for all $d$ measured 
\cite{Donev05a,Silbert06,Charbonneau12,Lerner13a}.

{\it Force distribution:} The distribution of the amplitude of contact forces $P(f)$ has been extensively characterized in granular materials \cite{Liu95,sno}, and its behavior at large forces has been a matter of intense debate. As we shall see,  the distribution at small forces has  more important physical implications.  There, $P(f)$ is  characterized by a non-trivial exponent \cite{Lerner12,Charbonneau12,Lerner13a}:
\be
\label{force}
P(f)\sim \frac{f^\theta}{\langle f\rangle^{\theta+1}}.
\ee  
For $d=2,3$ one finds   $\theta\approx 0.18$.
Most of the previous theoretical literature assumed that forces propagate on some frozen disorder  \cite{Liu95,Otto03,Bouchaud01}, which lead to the incorrect prediction that $\theta$ should be an integer.   

\subsection{Rearrangements under increasing stress}
Random close packings display crackling noise \cite{Combe00}. The controlling parameter is the anisotropy of stress $\mu\equiv F_2/F_1-1$, see illustration in Fig.~\ref{fig3}. As $\mu$ grows there are force intervals  where particles do not move at all,  due to their infinite stiffness. At discrete points in time, however, a contact force vanishes which leads to a rearrangement of the packing, and thus changes the contact network. The change of the box shape can be followed via the strain $\epsilon$, an intensive quantity defined in the caption of  Fig.~\ref{fig3}. Denoting $\Delta \epsilon$ the strain increment associated with these rearrangements, one finds that their distribution follows $P(\Delta \epsilon)\sim \Delta \epsilon^{\tau}$ with $\tau\approx -1.46$ for $d=2$. 


\subsection{Elementary Excitations}

\begin{figure}[ht]
\includegraphics[scale = 0.55]{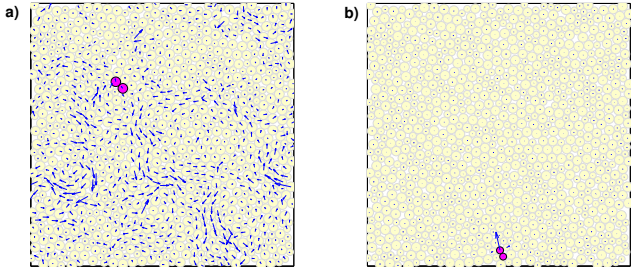}
\caption{
Examples of two  floppy modes, each obtained by pushing two particles apart in the same packing.  (a) Extended excitation and (b) localized buckling excitation.   }
\label{leverFig}
\end{figure}

In an isostatic system the removal of any contact leads to the creation of one floppy mode.
We consider as elementary excitations the floppy modes generated by opening  a contact labelled $\alpha$, while all the other contacts remain closed, see Fig.~\ref{leverFig}. We denote by $\delta {\vec R}^{(\alpha)}_{i}(s)$ the displacement of particle $i$ when $\alpha$ is opened by a distance $s$. This displacement field is uniquely defined, because only one floppy mode appears when a contact is broken, and exists for sufficiently small openings $s$ before new contacts are formed. Because contacts other than $\alpha$ remain closed one has:
\begin{equation}\label{foo02}
\delta r_{ij}=\delta\vec{R}^{\pair}_{\generalPair}(s)\cdot\vec{n}_{\generalPair} +
\frac{\left(\delta\vec{R}^{\pair}_{\generalPair}(s)\cdot \vec{n}^\perp_{\generalPair}\right)^2}{2r_\generalPair}+{\cal O}(s^3)
= s\delta_{\alpha,\generalPair}\ ,
\end{equation}
where $\delta_{\alpha,\generalPair} = 1$ if and only if the pair $\langle\generalPair\rangle$ is $\alpha$, and  zero otherwise, $\vec{n}_\generalPair \equiv \vec{R}_\generalPair/r_\generalPair$ is the director  pointing from particle $i$ to particle $j$, and $\delta\vec{R}^{\pair}_{\generalPair}\cdot \vec{n}^\perp_{\generalPair}$ indicates the projection of
$\delta\vec{R}^\pair_\generalPair$ onto the space orthogonal to $\vec{n}_\generalPair$. Eq.~(\ref{foo02}) simply stems from the Pythagoras theorem.

For a typical contact  in an isostatic system it can be shown that floppy modes are {\em extended}, i.e., $\delta {\vec R}^{(\alpha)}_{i}(s)\sim s$ \cite{Wyart05b}, as illustrated in Fig.~\ref{leverFig}a. 
Sometimes local displacements are however weakly coupled to the rest of the system, 
resulting in {\it local buckling modes}, cf., Fig.~\ref{leverFig}b.  
There is in fact a continuum of excitations between these two extremes cases, but as far as scaling is concerned it is sufficient to consider these extremes  \cite{Lerner13a}.
The local
buckling  modes dominate the force distribution at small $f$, 
$P_{\rm lb}(f \to 0)= P(f\to 0)\sim f^\theta/\langle f\rangle^{\theta+1}$, 
while extended excitations come with a force distribution $P_e(f)\sim f^{\theta'}/\langle f\rangle^{\theta'+1}$, where numerically one finds $\theta'\approx 0.44> \theta$ \cite{Lerner13a}. 

\subsection{Stability bound}
\begin{figure}[ht]
\includegraphics[scale = 0.4]{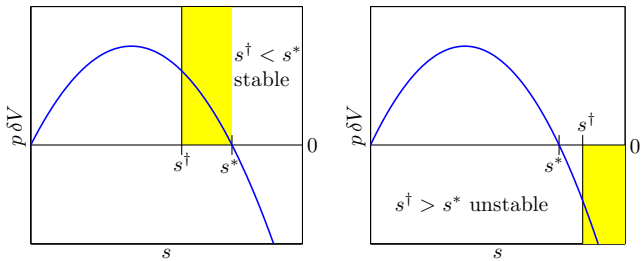}
\caption{ Volume change $p\delta V$ {\it vs} opening distance $s$ of a contact $\alpha$.
If the floppy mode forms a new contact at  $s^\dagger < s^*$ and thus stops,
the packing is stable. Otherwise, a \emph{denser} packing can be generated.}
\label{illustration}
\end{figure}

Let us now consider whether pushing two particles apart can lead to a denser packing and  thus release compression energy. The force balance in the unperturbed state can be written as:
\begin{equation}
\label{1}
\forall i, \ \  {\vec F}_i+\sum_{ j(i)} f_{ij} {\vec n}_{ij}=0\ ,
\end{equation}
where the sum is over all particles $j(i)$ in contact with $i$, ${\vec F}_i$ is the force exerted by the wall on particle $i$ (zero in the bulk), and $f_{ij}>0$ is the magnitude of the  repulsive) force in the contact $\langle ij \rangle$. Multiplying Eq.~(\ref{1}) by any displacement field $\delta {\vec R}_i$ and summing on all particles leads to the virtual work theorem:
\begin{equation}
\label{2}
\sum_i {\vec F}_i\cdot \delta {\vec R}_i+\sum_{\langle ij \rangle} \delta {\vec R}_{ij}\cdot{\vec n}_{ij} f_{ij}=0,
\end{equation}
where $\sum_{\langle ij \rangle}$ denotes the summation over all contacts $\langle ij \rangle$. 
The compression work done by forces from the boundaries is $\sum_i {\vec F}_i\cdot\delta {\vec R}_i=-p\delta \Omega$. 
Multiplying Eq.~(\ref{foo02}) by $f_{ij}$, and using Eq.~(\ref{2}), one obtains:
\begin{equation}\label{foo06}
p\delta\Omega^\pair = sf_\alpha - 
\sum_{\langle ij \rangle} \sFrac{\left(\delta\vec{R}^\pair_\generalPair\cdot \vec{n}^\perp_\generalPair\right)^2\! f_\generalPair}{2r_\generalPair} + {\cal O}(s^3)\ .
\end{equation}
Consider the stability of extended excitations for which $|\delta\vec{R}^\pair_\generalPair|\sim s$. Since there are of order $N$ terms in the sum, we can rewrite Eq.~(\ref{foo06}) as:
\begin{equation}\label{foo03}
p\delta \Omega = sf_\alpha - s^2c_\alpha N\langle f \rangle/\sigma_0 + {\cal O}(s^3)\ .
\end{equation}
where $c_\alpha=O(1)$ is a contact-dependent constant. 
For small opening $s$, the volume $\Omega$ always increases
since the force $f_\alpha$ is positive, as illustrated in Fig.~\ref{illustration}. However, for larger openings the nonlinear term in Eq.~(\ref{foo03})
becomes important, and for:
\begin{equation}
\label{111}
s > s^* \equiv \frac{f_\alpha \sigma_0}{c_\alpha\langle f\rangle N}\ ,
\end{equation}
a \emph{denser} packing can be generated. This condition is most constraining for the extended mode with the smallest contact force, $f_{\rm min}$,
satisfying $\int_0^{f_{\rm min}} P_e(f) df\sim \frac{1}{N}$, and thus $f_{\rm min}\sim \langle f\rangle N^{-1/(1+\theta')}$. Plugging this into Eq.~(\ref{111}) we find the threshold distance $s^*_{\rm min}$ beyond which the volume is reduced:
\be
\label{112}
s^*_{\rm min}\sim\sigma_0 N^{-1-1/(1+\theta')}.
\ee

However, $s$ is bounded by the displacement $s^\dagger$, where a first new contact is created. For extended modes, $s^\dagger$ is of order of the minimal gap $h_{\rm min}$ in the system, given by $\int_0^{h_{\rm min}}g(h')dh' \sim 1/N$, or  $s^\dagger \sim h_{\rm min}\sim \sigma_0N^{-\frac{1}{1-\gamma}}$.
Stability requires $ s^\dagger< s^*$,  cf. Fig.~\ref{illustration}, and for the mode with weakest force,  $s^\dagger<s^*_{\rm min}$ or  $\sigma_0N^{-\frac{1}{1-\gamma}}<\sigma_0 N^{-1-1/(1+\theta')}$. For large $N$, one thus has the bound
\be 
\label{gamma_bound}
\gamma\geq\frac{1}{2+\theta'} .
\ee
A similar reasoning for the localized modes leads to \cite{Lerner13a}:
\be
\label{10}
\gamma\geq\frac{1-\theta}{2}
\ee

Numerically both bounds are found to be marginally satisfied with $0.38\approx \gamma\geq(1-\theta)/2\approx 0.41$ and $0.38 \approx \gamma \ge 1/(2+\theta') \approx 0.41$ \cite{Lerner13a}, as 
one may expect from a dynamical picture as in Fig.~\ref{fig:marginal} \cite{Wyart12}.  A more rigorous proof is given in Sec.~\ref{ApplHS}.  

Marginal stability thus leads to a description of random packings based on three exponents constrained by two relationships. Very recently the exponent $\theta$ and $\gamma$ were computed in infinite dimensions using the replica method \cite{Charbonneau14,Charbonneau13}, leading to $\gamma=0.4126...$ and $\theta=0.4231...$ as defined from Eqs.~(\ref{gap},\ref{force}). Interestingly, these exponents exactly saturate the bound Eq.~(\ref{gamma_bound}), and not Eq.~(\ref{10}), presumably because localized buckling excitations are absent in infinite dimensions~\cite{DeGiuli14b}. It remains to be seen if the value for $\gamma$ is exact for $d<\infty$ as well, given that it is quite close to the numerical values in $d=2,3$.

\section{Effect of the pseudo-gap on dynamics}
\label{framework}

\subsection{Review of previous arguments for marginality}
In the existing literature on glasses, most arguments for the saturation of stability bounds concerned the equilibrium. Both for mean-field spin glasses~\cite{moore,thouless} and electron glasses~\cite{efros} it was originally argued that they have no particular reason to dig a stronger pseudo-gap than one consistent with the stability constraints, and thus one should expect saturated exponents. While this was later confirmed by an analytical calculation in the SK model~\cite{SommersDupont, Pankov}, in electron glasses it remains a conjecture at the level of dressed excitations in the equilibrium state, as we have discussed above. 

A theoretical justification for marginality in equilibrium is provided by replica mean-field theory, which is exact for the SK model, but approximate for electron glasses.~\cite{DobroPankov, markus, MarkusPankov} 
At the level of the effective potential approach~\cite{thouless} on finds that continuous replica symmetry breaking is equivalent to a gapless spectrum of the inverse susceptibility matrix, $\partial ^2F/\partial m_i\partial m_j$, implying the presence of arbitrarily soft collective modes involving a large number of spins. While this {\em thermodynamic} marginality has important consequences for equilibrium response, including equilibrium avalanches~\cite{MarkusEPL,MarkusSKavalanchesPRB}, it is less clear whether those modes can be excited by a single- or few-spin flip dynamics in the $T\to 0$ limit. Moreover, it is unclear whether equilibrium properties are at all relevant to understand properties in the out-of-equilibrium steady state configurations on the hysteresis loop. 
Even if the equilibrium states display marginality (e.g., in the form of continuous replica symmetry breaking) the corresponding out-of-equilibrium states might not share this property. This is suggested by the absence of crackling in short range spin glasses even in high dimensions $d \geq 8$ and on random regular graphs~\cite{Katzgraber} - both of which are believed to exhibit continuous replica symmetry breaking in equilibrium. In Sec.~\ref{SpinglassApp} we will explain that in both systems avalanches are bounded and become large only in the limit of infinite coordination number.

\subsection{Relation between equilibrium and driven steady state configurations}
 
From the above considerations one should expect that dynamically explored states differ quantitatively from more stable states, that would be reached by slow, long time annealing involving collective multi-particle relaxations. This is well illustrated by the SK spin glass, for which the analytical solution for the ground state yields the result $P(h)= 0.31 |h|$, while states on the hysteresis loop, and metastable states out-of-equilibrium are numerically seen to have a significantly larger slope of the linear pseudogap~\cite{Pazmandi,Parisi03,LeUs, Arielprivate}. This reflects the fact that those states are only stable with respect to a restricted set of spin flip operations, while the ground state is stable with respect to any number of spin flips, and thus has a stronger pseudogap suppression.

\begin{figure}[ht]
\includegraphics[scale = 0.6]{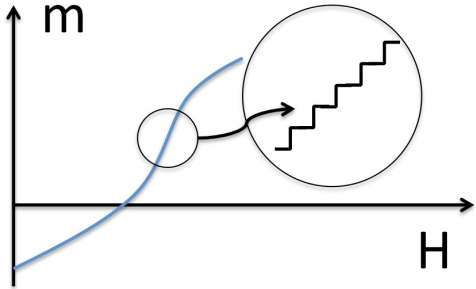}
\caption{Sketch of a portion of the hysteresis loop showing the magnetization per spin $m$ vs. the applied field $H$. This curve appears continuous for large system size, but consists in fact of discrete jumps. A qualitatively similar curve also describes hard sphere packings when plotting  strain $\epsilon$ vs.  stress anisotropy $\mu$, as well as Coulomb glasses (charge density vs. gate voltage $V_g$), soft amorphous solids (strain $\epsilon$ vs. applied shear stress  $\sigma$, or the depinning of an elastic interface (mean position of the interface $\langle y\rangle$ vs. applied force $F$). }
\label{hysteresis}
\end{figure}

\subsection{Pseudo-gap and avalanches}

We consider a glassy system at zero temperature, driven adiabatically by some field. For concreteness we choose spin glasses,
but as we will see below our arguments are much more general, cf.~Fig.~\ref{hysteresis}. As the field $H$ increases by some amount $\Delta H$ of order $O(1)$,
the magnetization per spin $m$ changes by $\Delta m = O(1)$ as well. Due to randomness this jump does not occur at a single value of the field, unlike in clean ferromagnets. We consider a one-spin flip dynamics. Hence, as long as all spins are aligned with their local field, increasing $H$ leads to no dynamics. These periods correspond to the horizontal lines in Fig.~\ref{hysteresis}. (In systems with continuous degrees of freedom these intervals exhibit a smooth but non-zero $dm/dh$. Our arguments apply as long as a finite fraction of the increase of $m$  occurs through jumps, which is guaranteed if the dissipation along a full hysteresis cycle is extensive.) However, once a  local field passes through zero and turns opposite to the spin, a dynamics of successive spin flips sets in (at fixed $H$), which lasts until all the spins are aligned with their local fields again. These rearrangements correspond to the discrete vertical jumps in the hysteresis curve in Fig.~\ref{hysteresis}.

Above we discussed stability with respect to two simultaneous events (flipping two  spins in the spin glass, emptying one state and filling another in the Coulomb glass, or opening one contact and closing another in packings).
Let us now explain why the exponent bounds are still satisfied, 
even if one only considers the more restrictive one spin-flip moves. To this end we introduce the quantity $E$, also used in depinning problems \cite{Bak87}, which characterizes the average number of spins which become unstable when one unstable spin flips ~\footnote[1]{Note that this average  can depend on the instantaneous number of unstable spins, or other variables. $E$ is averaged over these variables.}. If $E\gg 1$, the number of unstable spins explodes exponentially with time and a huge avalanche occurs. By contrast, if $E\ll 1$, one spin flip very rarely triggers another one, and rearrangements consist of one, or very few, spin flips. The intermediate case of power-law distributed avalanches are thus possible only if $E\approx 1$.

If the stability bound on the exponent is violated, the number of unstable pairs of spins diverges in the thermodynamic limit. This turns out to be equivalent to the divergence of the quantity $E$, both for power-law or mean-field interactions. For mean-field spin glasses for example, if $P(h)\sim h^\theta$, a spin flip induces of the order of $E\sim N \int_0^{J_{\rm typ}\sim  1/\sqrt{N}} P(h) dh\sim N^{(1-\theta)/2}$ further spin flips, which indeed diverges if $\theta<1$. Since configurations with $E\gg 1$ cannot be  the endpoint of an avalanche with one spin-flip dynamics, the two spin-flip stability bound on $\theta$ must be satisfied for that dynamics as well. Note that while this consideration constrains the exponent, it does not guarantee the absence of any unstable pairs of spins after the avalanche has stopped.
Nonetheless, it is unlikely that many unstable pairs survive, as it is probable that they are either eliminated, or not created in the first place, by sequential spin flips in an avalanche. This is the reason why one finds well developed pseudogaps even in metastable states for which only single spin stability is imposed.~\cite{HFMarkus, Palassiniunpublished}

In the case where the interaction is mean-field, as in the SK model or packings of hard spheres, the argument can be carried further. If the exponent bound were not saturated, $E$ would vanish in the thermodynamic limit. This is not true for power-law interactions where short-range interactions always lead to a finite $E$. Thus, in a mean-field situation, if the bound were not saturated, rearrangements would  essentially occur by single spin flips. As we shall see now, this would be contradictory, the only consistent solution being  the marginal exponent. 

\subsection{Pseudo-gap implies large mean avalanche size}
We now argue that  the presence of a pseudo-gap {\it implies} large avalanches. This is straightforward to show, but has ample consequences. Assuming a pseudo-gap $P(h)\sim h^\theta$, the smallest field in the sample is easily found to scale as
 $h_{\rm min}\sim N^{-1/(1+\theta)}\gg 1/N$. Thus, the number of rearrangements $N_a$ triggered as the field is swept by $\Delta H\sim1$ is of order:
\be
\label{numbera}
N_a\sim 1/h_{\rm min}\sim N^{1/(1+\theta)}\ll N,
\ee
and hence much smaller than $N$. However, taken together these rearrangements must change the total magnetization extensively, implying that the mean size of rearrangement events, $\langle M_a\rangle $, diverges in the thermodynamic limit as
\be
\label{mean}
\langle M_a\rangle\sim N/N_a\sim N h_{\rm min}\sim N^{\theta/(1+\theta)}.
\ee
Such large rearrangements can only occur if $E\rightarrow 1$ as $N\rightarrow \infty$, which leads to broadly distributed avalanches with a cut-off diverging with $N$. 
Thus crackling must occur as soon as a pseudo-gap is present, i.e., whenever $\theta>0$. This is a central result, as it shows that long-range interactions are relevant as far as crackling is concerned: since they necessarily open a pseudo-gap, they must always lead to  avalanches whose mean  diverges with $N$. 

\subsection{Mean-field systems are marginally stable}
We have argued that mean-field interactions imply a pseudo-gap, that a pseudo-gap implies crackling, and that in mean-field models crackling can occur only if stability is marginal. This argument thus proves that within the hysteresis loop of the SK model the stability must be marginal, and the exponent  in Eq.~(\ref{spinglassbound}) is  saturated. As we will see, the same result holds for hard spheres. 

\subsection{Purely short-range interactions do not crackle on the hysteresis loop}
In systems with frustrated short-range interactions, as discussed above, one expects no pseudo-gap, i.e., $P(h=0)>0$ (at intermediate fields $H$). Therefore, $h_{\rm min}\sim 1/N$,
and from Eq.~(\ref{mean}) one obtains $\langle M_a\rangle\sim 1$: the mean jump of magnetization during rearrangements does not diverge in the thermodynamic limit. Strictly speaking, this does not imply the impossibility of power law-distributed avalanches with a cut-off $S_c$ that diverges as $N\rightarrow \infty$.
They could occur, but only with an unusual size distribution $\rho(S)\sim S^{-\tau}$ with $\tau>2$. We are not aware of simple models where this condition is satisfied. Moreover, we argue below that this condition does not arise in short-range spin glasses in particular, where we expect $\tau\leq 3/2$ for $S$ below some finite $S_c$. We believe that this situation is generic, cf. also Ref.~\cite{Katzgraber}: No crackling occurs on the hysteresis loop of models with short-range interactions, except at specially fine-tuned values of couplings and fields~\cite{vives}.

\section{Applications}

\subsection{Spin glasses}
\label{SpinglassApp}

{\it Fully connected spin glasses (SK model):}\\
For the SK model the above considerations confirm previous results in the literature~\cite{Pazmandi}: A linear pseudo-gap is established along the hysteresis loop. Accordingly scale free avalanches occur in intervals of $dH\sim 1/\sqrt{N}$,  with average magnetization jumps $dM \sim\sqrt N$.  \\

{\it Short-range spin glasses:}\\
It is believed that the thermodynamics of spin glasses on finite-dimensional lattices or random graphs tends to the mean-field limit captured by the SK model, as the coordination number $Z$ becomes large. However, with respect to dynamics, and avalanches in particular, such a correspondence at large $Z$ is in fact not to be expected. The SK model has the property that any two sites which are neighbors of a given site, are also neighbors of each other, very much in contrast to high-dimensional lattices and random graphs, where this is very rarely the case.  This example illustrates the crucial difference between thermodynamics and out-of-equilibrium dynamics.

In the SK model an avalanche progresses due to sign flips of local fields that are brought about  collectively by all spins that have flipped previously. In contrast, avalanches on random graphs consist simply of a local tree-like structure of flipping spins, where every spin flip is triggered only by its unique ancestor in the tree. We expect avalanches in high dimensional lattices to be very similar in character, due to the small likelihood of mutual crossings of different avalanche branches. Neglecting the potential field correlations that may establish during the hysteresis cycle, such avalanches are simple realizations of a Galton-Watson type branching process with a certain average branching ratio $E$. On the hysteresis loop it will tend to a constant rather close to, but smaller than $1$. 
Accordingly one expects avalanche sizes to be distributed as $P(S)\sim S^{-\tau} \exp[-S/S^*]$ with $\tau =3/2$ and finite cut-off $S^*$. The average size of avalanches thus follows as $\langle S\rangle =1/(1-E)\sim \sqrt{S^*}$. 

We can determine $S^*$ (and thus $E$) by considering a compact set of $N\gg 1$ spins. The minimal local field in this set scales as $h_{\rm min} = [N P(0)]^{-1}\sim \sqrt{Z}/N$. Along the main part of the hysteresis cycle there will thus be $\sim 1/h_{\rm min}$ avalanche events affecting this set. Each spin flip in an avalanche typically dissipates an energy of order $J_{\rm typ}\sim 1/\sqrt{Z}$. The total energy dissipated along the hysteresis curve  thus amounts to $E_{\rm dis}\sim h^{-1}_{\rm min} \langle S\rangle J_{\rm typ}$, or $\langle S\rangle/Z$ per spin. Note that $E_{\rm dis}$ corresponds to the width of the hysteresis curve (per spin), which we expect to have a finite limit as $Z\to \infty$.
We therefore obtain the predictions $\langle S\rangle \sim Z$ and $S^*\sim Z^2$. 
This theoretical consideration is in good agreement with the numerical results on high-dimensional lattices, especially with the empiric finding $S^*\sim Z^2$.~\cite{Katzgraber}  \\

{\it Mixing long and short-range interactions:}\\
In order to illustrate the relevance of long-range interactions, it is instructive to  consider the Hamiltonian~\cite{marco}:
\be
\label{mix}
H=-\frac {1}{2} \sum_{i\neq j}  s_i J_{ij} s_j  - \frac {1}{2} \sum_{\langle i, j\rangle}  s_i J'_{ij} s_j,
\ee
where ${\langle i, j\rangle}$ indicates a pair of neighboring particles on a lattice of coordination $Z$, and $ \overline{J' {}^2_{\langle ij\rangle}}=1/Z$.
The first term is of infinite range with $ \overline{J_{ij}^2}=\tilde J^2/N$. When $\tilde J=0$, Eq.~(\ref{mix}) 
describes an Edwards-Anderson spin glass, for which $P(0)\approx 1/\sqrt{Z}$  as argued above. However, for $1\gg \tilde J>0$, stability requires a pseudo-gap
to form. From the arguments of Sec.~\ref{spin} one concludes that $P(h)\leq a_0 |h|/\tilde J^2$, where $a_0$ is some constant of order $O(1)$.
Using Eq.~(\ref{mean}) we find that the mean jump of magnetization in the hysteresis curve is of order $\langle M_a\rangle \sim \tilde J \sqrt{N}$.
Thus, crackling occurs even if $\tilde J \ll 1$, although it is absent for $\tilde J=0$. \\

\subsection{Electron glasses}
In an electron glass in 2d or 3d the above considerations prove that there must be scale-free avalanches if the dynamics allows the addition or removal of a particle on single sites, as would be the case for a system coupled everywhere to a bath of particles. Indeed, in that case the relevant excitations are charge insertions and removals, described by the single particle density of states $\rho(E)\sim |E|^{d-1}$ on the hysteresis loop. 

However, if there is no such particle bath, as in a sample coupled to leads only at the boundaries, the relevant excitations are compact dipolar particle-hole excitations, whose interaction decays as $1/r^3$. Stability arguments on those excitations imply no pseudo-gap in $d=2$ and only a logarithmic suppression in $d=3$. The size of avalanches triggered by dipole excitations  thus grows at best logarithmically with system size (under the condition that the compactness of dipolar excitations is enforced), consistent with the numerical observations in Ref.~\cite{PalassiniGoethe}.

\subsection{Hard spheres}
\label{ApplHS}
Hard spheres behave similarly to  fully-connected spin glasses, because the effective interaction (mediated by elasticity) is not decaying with distance. In particular it was shown  \cite{Wyart12,Lerner13} that if the bound Eq.~(\ref{gamma_bound}) is not satisfied, extensive avalanches take place. If instead it is strictly satisfied, contacts open one by one as the stress anisotropy increases. To prove marginal stability we thus simply need to show that the average number of contacts opening during rearrangements diverges with $N$. As will be shown elsewhere \cite{Lerner14bis} the characteristic increment of stress anisotropy $\delta \mu$ needed to open one contact follows $\delta \mu\sim N^{-1/2-1/(1+\theta')}$ (extended contacts open first), while the jump of strain resulting from opening a single contact scales as $\delta \epsilon = N^{-3/2-1/(1+\theta')}$, in agreement with numerics \cite{Combe00}. Assuming that the total variation of both $\epsilon$ and $\mu$ is $O(1)$ along the hysteresis loop, the mean change of strain in one avalanche is then $\langle \epsilon_a\rangle \sim 1/\delta \mu \sim N\delta \epsilon \gg\delta \epsilon$, where the last inequality implies crackling noise.

\subsection{Depinning of an elastic interface}
We  now consider the depinning transition of an elastic interface of dimension $d$ embedded in a space of dimension $d+1$ \cite{Fisher98, Kardar98}. When disorder is present, there is a critical force $F_c$ per unit surface beyond which the manifold is not pinned, but moves with some average velocity $v$. At $F=F_c$ where $v$ vanishes the dynamics becomes very jerky~\cite{Fisher98, Kardar98}. For $F<F_c$ the manifold is pinned, and  avalanches occur as the force $F$ is increased adiabatically. Avalanches are power-law distributed with a cut-off which diverges only as $F\rightarrow F_c$. We argue that this last fact is consistent with our arguments, which predict no crackling if the manifold is driven strictly below $F_c$ at $T=0$. 

For an elastic interface, local elementary excitations can be defined as follows. Cut the interface into small cells and define the additional local force $\delta F_i$ necessary to induce a local depinning event where the cell $i$ exits the pinning potential that traps it. The probability distribution $P(\delta F)$ characterizes the density of excitations. It is well known that $P(\delta F=0)>0$ \cite{Fisher98}, if the elastic interaction characterizing the interface is {\it monotonic }\cite{Lin14}, which means that a yielding cell always tends to destabilize other cells, but never stabilizes them. This implies that locally the distance to  instability $\delta F_i$ always decreases as $F$ increases, until $\delta F_i<0$ at which point the cell $i$ rearranges. Nothing in the dynamics allows the cell $i$ to forecast that an instability approaches, and hence, no depletion nor accumulation can occur near $\delta F_i=0$. 

The argument developped for spin glasses in the $m$ {\it vs} $H$ plane then easily translates to the problem of depinning. One considers the evolution of the mean position of the interface $\langle y\rangle$ as a function of $F$. This is well-defined and independent of time as long as $F<F_c$, the $F$-dependence being qualitatively similar to Fig.~\ref{hysteresis}.   Since $P(0)>0$, $\delta F_{\rm min}\sim 1/N$, and Eq.~(\ref{mean})  implies that the mean of the rearrangements of the interface is independent of the system size: no crackling occurs for $F$ sufficiently below $F_c$~\cite{Middleton}. As in spin glasses, this differs from equilibrium avalanches, for which analytical studies find power laws with a cut-off that diverges as a regularizing mass tends to zero.~\cite{staticavalanches_manifold,MarkusEPL}

\subsection{Plasticity in amorphous solids with soft interactions}

We now consider soft amorphous solids, such as foams, emulsions, or metallic glasses, in conditions where thermal fluctuations can be neglected.  If the applied shear stress $\sigma$ is increased adiabatically, the elastic energy loads continuously, and is released by sudden local plastic events,  the so-called shear-transformations \cite{Argon79,Falk98}.  These elementary excitations are elastically coupled with each other \cite{Hebraud98}, and one plastic event can trigger many others, potentially leading to avalanches. The interaction is long range (it decays in amplitude as $1/r^d$ \cite{Picard04, Nicolas14})  and non-monotonic: a plastic event can either stabilize or destabilize other regions, depending on their relative location.  

Let us again cut the amorphous solid into small  blocks containing several particles and define as $\delta \sigma_i$ the additional stress to be applied to the block  $i$ to induce an instability.  The probability distribution $P(\delta \sigma)$ is a measure of how many putative shear transformations are present in the sample \cite{Lin14,Lin14b,Karmakar10a}. Due to the long-range interaction and the randomness of signs, one finds a stability criterion essentially identical to that of dipolar glasses discussed above. More precisely, $P(\delta \sigma)$ must vanish logarithmically as $\delta \sigma \rightarrow 0$ ($\theta=0$), otherwise extensive avalanches would occur~\cite{Lin14}. 
Numerically, this constraint is not saturated however. Rather, the pseudo-gap is stronger, with $\theta\approx 0.35$ in $d=3$ and $\theta\approx 0.55$ in $d=2$ as found in automaton models \cite{Lin14,Lin14b}, in agreement with finite size effects observed using molecular dynamics \cite{Karmakar10a,Salerno12,Salerno13,Maloney04}. 

The exponent $\theta$ enters the scaling description of the yielding transition that occurs beyond some yield stress $\sigma_c$ \cite{Lin14b}. 
Here instead we focus on the hysteresis occurring for $\sigma<\sigma_c$, schematically represented in Fig.~\ref{hysteresis} with $x$- and $y$-axis corresponding to the stress $\sigma$ and the  strain $\epsilon$, respectively. 
By contrast with the depinning transition, our arguments predict that crackling occurs within the hysteresis loop, with avalanches of mean size $M_a\sim N \delta \sigma_{\rm min} \sim N^{\theta/(1+\theta)}$. Numerical observations of such crackling in models of plasticity in solids below the yield stress were reported very recently~\cite{ispanovity13}.

\section{Discussion}

{\it Dynamical interpretation of the role of long-range interactions:}  Above we have argued that glasses with long-range interactions feature a pseudo-gap, which implies avalanches with diverging sizes. This in turn implies, that the expectation value $E$ of spin flips rendered unstable by a given spin flip is very close to $1$, up to a negative correction which vanishes as an inverse power of $N$. It is instructive to discuss how this expectation value becomes pinned so precisely in the case of power law interactions. Consider the flip of spin $s_0$ within an avalanche. First, the nearby spins, which are more strongly coupled to $s_0$, adjust to the flipped $s_0$ and relax. Subsequently more energy can be relaxed by adjusting more  distant spins with weaker coupling. The process only stops as the system size is reached. While the nearby spins are flipped the local value of $E$ (defined for flips of spins in the vicinity of $s_0$) is decreasing in discrete steps. However, as more and more distant spins are relaxing, $E$ depends just on the evolution of $P(h)$ characterizing the large set of spins at a great distance. Since $P(h)$ evolves essentially continuously during an avalanche (being an average over many spins), $E$ decreases in ever finer steps, which scale with the inverse system size. The avalanche stops as soon as  $E$ is  smaller than $1$, and thus it is natural that $1-E$ will be small as an inverse power of $N$. In contrast, short range systems have no similar reason to settle into states with such fine-tuned $E$, which is why their avalanches remain bounded. 

{\it Saturation of stability bound:} In many cases of long-range interactions it has been found that the stability bounds on the exponents are saturated in the quasi-adiabatically driven states. We expect that it is also the case for the model of Eq.~(\ref{mix}). However, there are also some notable exceptions, e.g., in plasticity where the spatial correlated nature of avalanche events seems to suppress pseudo-gaps beyond the naive bound based on two-particle stability, an important point to clarify in the future.  
 
{\it Avalanches statistics:}  the exponent $\tau$ remain a challenge to understand in a number of cases. It is interesting that in mean-field spin glasses the difference between the equilibrium and the dynamic responses is much less dramatic than for short range models. This is presumably related to the fact that dynamics in the SK model is rather different from that in short range glasses, even if their connectivity is high. The fact that in the SK model many spin flips work together to make further spins flip is probably a crucial element in bringing about a behavior much closer to equilibrium jumps, where highly collective rearrangements play a central role. A better theoretical understanding of the avalanche dynamics in the SK model is however still highly desirable, see Refs.~\cite{moore,horner} for some attempts in this direction. Further work on this issue is in progress.~\cite{LeUs}
 

%
 \section{Conclusion}
Some physical systems sit at a critical point and display crackling noise when they are slowly driven. Such self-organized criticality occurs for example for an elastic manifold moving in a random environment. Here we have argued that a distinction should be made between these situations, and the marginal stability that characterizes glassy materials with long-range interactions. In these materials, crackling noise occurs because the configurations that are accessible dynamically after a rapid quench are barely stable and exhibit a pseudo-gap. Since this property holds independently of the field amplitude, crackling noise occurs for a range of fields, unlike for the depinning transition. 

 In this review we have focused on glasses. It remains to be seen if marginal stability extends to complex dynamical systems such as those that pervade biology or economics. Evolving species or economic agents for example can interact with many of their peers, often with antagonistic goals. It would thus be interesting to see if the notions of a marginal manifold, elementary excitations and pseudo-gaps are useful to describe these systems and their dynamics as well.

\section{Acknowledgments}
We thank Jie Lin, Le Yan, Edan Lerner, Gustavo During, Eric DeGiuli, Marco Baity Jesy, Alberto Rosso, Pierre Le Doussal and Matteo Palassini for useful discussions. MW thanks the National Science Foundation (NSF) CBETS Division Grant 1236378, NSF DMR Grant 1105387 and the MRSEC Program of the NSF, Grant DMR-0820341 for  support.

\end{document}